\documentclass{elsart}

\usepackage{mathptmx}
\usepackage{amsmath, amscd, amssymb}
\usepackage{mathrsfs}

\journal{Physica A}

\newcommand{\FT}{\Phi^\times}
\newcommand{\F}{{\Phi}}

\newcommand{\ts}{\tilde\psi}
\newcommand{\la}{\lambda}

\newcommand{\G}{\Gamma}

\newcommand{\HH}{{\mathscr{H}}}

\newcommand{\rhs}{\F\subset\HH\subset\F^\times}

\newcommand{\Om}{\Omega}
\newcommand{\om}{\omega}

\newcommand{\R}{{\mathbb R}}
\newcommand{\RR}{{\mathbb R}}
   
\newcommand{\II}{{\mathbb I}}
\newcommand{\C}{{\mathbb C}}

\newcommand{\CG}{{\mathscr{G}}}

\newcommand{\CS}{{\mathscr{S}}}

\newcommand{\g}{{\mathfrak{g}}}

\newcommand{\bee}{\begin{enumerate}}
\newcommand{\ene}{\end{enumerate}}
\newcommand{\een}{\end{enumerate}}
\newcommand{\bea}{\begin{eqnarray}}
\newcommand{\eea}{\end{eqnarray}}
\newcommand{\beq}{\protect{\begin{equation}}}
\newcommand{\eeq}{\protect{\end{equation}}}

\begin{document}

\begin{frontmatter}

\title{{Generalized Wavefunctions for \\
Correlated Quantum Oscillators II:\\
Geometry of the Space of States.}}

\author{S. Maxson}
\address{Department of Physics\\University of Colorado
Denver\\Denver, Colorado 80217\\ }
\ead{steven.maxson@ucdenver.edu}

\begin{abstract}

In this second in a series of four articles we create a mathematical formalism 
sufficient to represent nontrivial hamiltonian quantum dynamics, including resonances.
Some parts of this construction are also mathematically necessary. The specific
construction is the transforming of a pair of quantized free oscillators
into a resonant system of coupled
oscillators by analytic continuation which is 
performed algebraically by the group of complex symplectic
transformations, thereby creating dynamical representations of numerous 
semi-groups.  The quantum free oscillators 
are the quantum analogue of classical action angle variable solutions for the 
coupled oscillators and quantum resonances, including Breit-Wigner 
resonances.  Among the exponentially decaying Breit-Wigner resonances 
represented by Gamow vectors
are hamiltonian systems in which energy transfers from one oscillator 
to the other.  There are significant mathematical constraints in order 
that complex spectra be accommodated in a well defined formalism which
represents dynamics, and these
may be met by using the commutative real algebra $\C (1,i)$ as the
ring of  scalars in place of the field of complex numbers. These
mathematical constraints compel use of fundamental (spinor) 
representations rather than UIR's. By 
including distributional solutions to the Schr{\"o}dinger equation,
placing us in a rigged Hilbert space, and by using the 
Hamiltonian as a generator of canonical transformation of the space 
of states, the Schr{\"o}dinger equation is the equation for 
parallel transport of generalized energy eigenvectors, explicitly 
establishing the Hamiltonian as the generator of dynamical
time translations in this formalism.  

\end{abstract}


\end{frontmatter}

\section{Introduction}\label{sec:intro}

\subsection{Motivation}\label{sec:motivation}

The object of this second installment of a four part series is to
illustrate our method for constructing the most general possible
probabilistic description of dynamics which has the most well defined
mathematical structures possible.  We have indicated the general
mathematical structures and basic notions of a probability amplitude
description of correlated dynamics in installment
one~\cite{I}.  We will use the variable names $p$ and $q$ herein 
since they represent canonical variables.  We
will also understand the creation and destruction operators in a
slightly different sense than is traditional: we will think of $A$ and
$A^\dag$ in the first instance as vectors generating displacements in
coordinate directions in phase space (and its complexification)
when viewed classically and as the corresponding
operators in our function space representation of correlated dynamics
by probability amplitudes. (They are clearly the generators of translations,
but not infinitesimal generators. Hence, their identification as
vectors. There will be much more justification for this in installment
four of this series~\cite{IV}.) Since we will be working in the rigged Hilbert
space, $A$ and $A^\dag$ are not dual to each other as they were in the 
conventional Hilbert space approach: we will need another dual operation
than the complex conjugation of the Hilbert space version of quantum theory.
Therefore, $A$ and $A^\dag$ are merely a pair of operators for us, both
abstractly and in their function space representatives. We 
mention such things at the outset to give a taste of just how different
things must be in order to bring a large number of mathematical structures
into compatibility. We will see interesting physics emerge from the
unfamiliar mathematics.

There are many lessons on many levels which emerge
when we adopt the attitude that, if classical hamiltonian dynamics is
grounded in the study of 
symplectic structures and symplectic transformations
on phase space~\cite{souriau}, then quantum dynamics should be
based on the study of symplectic structures and symplectic
transformations associated with the
quantum mechanical space(s) of states.  We will adopt such an attitude
herein, and we will discover it takes very specific mathematical
formalisms to insure our efforts are as well defined as it is possible
to make them: be forewarned that  
unitary transformations are a subgroup of the symplectic
transformations, and hermitean operators are a special (i.e.,
symmetric) case of the essentially self-adjoint operators, and we will
adopt the more general types of transformations to our uses, 
thereby violating an orthodoxy entirely appropriate in another
context.  We will not be in von Neumann's Hilbert space anymore!  (Of
course, the very act of analytic continuation is a non-unitary
transformation, so the introduction of correlation notions by way of
analytic continuation into any Hilbert space description takes you
from the Hilbert space anyway~\cite{gadella}. The reader is reminded that 
the unit imaginary, $i$, is a type of correlation map.)

This approach also presents many threads to many notions currently
under investigation in other contexts.  Thus, if decoherence is the 
conversion of quantum probability
into a classical probability, we showed in the first installment that
our quantum construction can be thought of as
the analytic continuation (complex symplectic transformations) of a
type of classical probability theory.   
In this view, there should naturally be complex symplectic 
transformations which will have the effect of realification of our
(spinorial) wave functions.  We thus present the view that quantum
theory is the result of the addition of coherence (e.g., correlation)
to a
particular form of classical probability theory, and the complimentary
procedure of ``decoherence'' is also within the mathematical scope of
the construction.  In the third installment of this series, we will study
many quantum 
analogues to classical dynamical systems, and their relation to
classical statistical mechanics and thermodynamics. 

From the combination of the seemingly simple problem of canonical
transformations of two harmonic
oscillators, and empowered by the mathematical tools of generalized
functions (distributions) 
in a RHS, an astonishingly rich structure emerges
exhibiting many of the features one associates with classical hard chaos 
and irreversible classical thermodynamics (in the $N=2$ limit).  The
methods used here have transparent generalizations to larger numbers
of oscillators.  

The constructions of this paper make use of five major elements which
distinguish it from the conventional Hilbert space treatment of
quantum systems:
\begin{enumerate}
\item{It includes the weak, or distributional, solutions to the
  Schr{\"o}dinger equation, mentioned above.  These were not available
  to von Neumann.}
\item{For our ring of scalars, rather than choosing the unit imaginary,
  $i$, as an element in the field of complex numbers, we will use $i$
  as a $90$ degree rotation in the complex plane, i.e., necessarily as
  an element of a real algebra.  This means that for us $i$ is an
  element of the real algebra with two units, $\C (1,i)$. 
  The result is a complex hyperbolic
  (Lobachevsky) structure in the tangent space at every point in our
  complex spaces of generalized states (e.g., tangent bundle).We are
  building spaces (modules) whose ring of scalars are real algebras
  rather than spaces (modules) over fields.  This is the result of  a
  uniqueness consideration, based on the old 
  saw that complex conjugation is not a uniquely defined involution of
  the field of complex numbers regarded as an algebra--there is more
  than one possible conjugate structure to be obtained from the field
  $\C$.  A linear space is an algebra plus a scalar product. If $\F
  \subset \FT$, then the dual (adjoint) operation works an involution
  of $\F$ when considered as a linear space in this sense.  There is also 
  the canonical inclusion $\g \subset \g^\times$, indicating that there is an
  algebra involution associated with the dual operation applied to an
  operator Lie algebra. This also
  distinguishes the present work from other recent work using the RHS
  formalism.}
\item{We shall adopt an algebraic approach, emphasizing the 
  involution of real algebras, so that, e.g., the dual operation will
  be an involution of some sort. We will wish ultimately to identify
  the algebra of observables with the Lie algebra of dynamical
  transformations, which is the Lie algabra of the symplectic group
  of our phase space. This dual as involution then effects the
  canonical inclusion of the Lie algabra into its dual Lie algebra.}
\item{We will use the {\em weak} symplectic structure available on
  $\F\times\F^\times$ (e.g., associated with the scalar product on
  $\F$) to reflect (represent) the canonical weak symplectic structure
  on $\g \times\g^\times$ for Lie algebra $\g$ (which is represented
  on $\F$).  For the complex Hilbert space of von Neumann $\HH$, there
  is a {\em strong} symplectic structure on $\HH\times\HH^\times$,
  e.g., associated the scalar product on $\HH$~\cite{note1}. You
  cannot have a non-trivial dynamics in an Euclidean geometry.}
\item{In order to proceed uniquely, we are required to work with the
  real form (symplectic form) of various complex Lie algebras (of
  complex simple Lie groups), and representations of those complex Lie
  algebras.   This will operate as a major determinant of structure
  for our spaces of states (which are, due to this mathematical
  necessity, spin spaces, and our representations will be fundamental 
  representations and not unitary irreducible representations).}
\end{enumerate}

The basic program underlying the present work is conceptually quite
austere, although there are a myriad of details to the actual implementation
as we seek to implement compatibility of many disparate mathematical
structures. We regard $i$ as, e.g.,  an element of the commutative 
real algebra $\C (1,i) \equiv \R \oplus i\circ\R$~\cite{note2}.
We will regard the Hamiltonian of our oscillator system as a
realization of one of the generators of infinitesimal canonical 
transformations of the associated phase space. This makes 
it a generator in the Lie algebra of the appropriate symplectic 
(semi-)group representation,  
e.g., for two pairs of canonical coordinates, $H$ belongs to 
$\mathfrak{sp}(4,\R )^\C$.  Since $Sp(4,\C )$ is simple, it's
representation provides a connected covering structure in which 
to exponentiate our continuous infinitesimal symplectic transvections.  In fact,
the 
Schr{\"o}dinger equation would emerge as the analogue of Hamilton's
equation if we should chose a variational (i.e., Hamilton-Jacobi) 
treatment mathematically alternative to the approach adopted here.  

There is a significant incompleteness to our work at the present stage, as
convergence of the exponential map will not be addressed until
the fourth part of our series of papers~\cite{IV}. There we will see the
the application of an extraordinary result for spinors: conjugation by a
group element, $gAg^{-1}$ can be uniquely identified with a transformation
from the left only from the relevant spin group: $gAg^{-1}=h_{spin}A$. The
conjugation is well defined only in a finite local neighborhood (coordinate 
patch) for our semiigroups, but the
spin transformation which is thus fixed has global import. This admits an
extension to the semigroups of interest to us, and the
multicomponent nature of the vectors which is necessary for other reasons
in this part of our work will become crucial in another context when we 
address the exponential map, the spin nature of our constructions, and
related topological, group, semigroup and other issues in part four.

Our
Hamiltonian (and its Lie algebraic dual) can also be defined 
to have an infinitesimal action on the spaces of states in such a way
that its action on the space(s) of states is symplectic as well.  This
will ultimately result in it (the representation of the Hamiltonian) being 
associated with generating flows of hamiltonian (integrable) vector fields
on our generalized spaces of states (which are, in turn, representations of 
vector fields on phase space. 

We are creating a forum in which
we may speak of a quantum dynamics evolving through dynamical
(=symplectic) transformations in a manner which follows much of the
spirit of the classical treatment of dynamical systems.  
(The ``quantum chaos'' associated with the flow of topologically
transitive hyperbolic affine transformations on the spaces of quantum
states, which also have a symplectic action on the spaces of states,
will be discussed in the third installment of this series~\cite{III}.)
These mathematical structures are not compatible with the conventional  
formulation of quantum mechanics using von Neumann's Hilbert space,
and so we are pursuing a description of a ``quantum dynamical system''
which is unique to our variant of the RHS formulation of quantum
mechanics.  In the course of these investigation of formulation of a
quantum dynamics, many insights emerge illuminating mathematical
structures that naturally arise when one formulates an {\em
hamiltonian} quantum field theory based on
harmonic oscillators dynamically evolving in our version of
the RHS formalism.  There is a lot of mathematical physics we will
show some consistent
connection to which is above and beyond the machinery we
actually invoke for our immediate purposes.  (Roger Newton dealt with
enlarging the physical Hilbert space in order to obtain quantum
action-angle variables using ``spin like'' multi-valued quantum
numbers~\cite{newton}.  We are using spinors explicitly.  See
installment four~\cite{IV}.)  

Every complex semi-simple Lie algebra has some
complex Lie group, and on a semisimple Lie group with a complex Lie algebra, 
$exp$ is holomorphic~\cite{knapp}.  In consequence, the complex
simple Lie (semi-)groups relevant to our problem are connected and locally
path connected.  Since $exp$ is holomorphic for
$\mathfrak{sp}(4,\R )^\C$, the exponential map of $H$ can locally be
identified with the transformation of 
parallel transport along geodesics in $Sp(4,\R )^\C_\pm$.
Formally, we work with the real (or symplectic) 
form of the Lie algebra, e.g., $\mathfrak{sp}(4,\R )^\C$ rather 
than $\mathfrak{sp}(4,\C )$ itself, in order
that adjoints be well defined for both the algebra and the group
simultaneously.  (See Section~\ref{sec:scattering}.)  We will
construct a representation of this Lie group/Lie algebra structure,
and, with the only mathematical limitation being the assumption that
the potentials in the Hamiltonian are analytic, 
we may make use of the creation and destruction operator formalism.
(See~\cite{barut}.)  There are however some wrinkles, since our
distributional solutions are associated with semigroups rather
than full groups, and our geodesics may have only local uniqueness due
do topological considerations which will become apparent in installment
four~\cite{IV} - this is a local quantum theory in many senses. (We will eventually elaborate
in installment four~\cite{IV} how the exponential map may be holomorphic
and also how the semigroups of unitary transformations--viewed as a 
sub-semigroup of the symplectic semigroup--differ at most on a set of 
measure zero from the related unitary group.)

We construct our representation
spaces as modules over $\C (1,i)$, and corresponding to the
geodesics in $Sp(4,\R )^\C_\pm$ generated by $exp( \mathfrak{sp} (4,\R
)^\C )$ there will be geodesics of evolution
generated in our representation space by the representation of the Lie
group and associated Lie algebra: among the equations of parallel
transport along these geodesics, it is possible to find an equation equivalent 
to the Schr{\"o}dinger equation. 
We will (necessarily) allow weak solutions to these equations of
geodesic evolution on our representation space, so that our chosen
representation spaces include spaces of generalized functions, and thereby
our work naturally falls into the RHS formalism.  As part of our
construction, we endeavor to respect all canonical constructions,
including canonical inclusions (Section~\ref{sec:scattering}) and
canonical symplectic forms (Section~\ref{sec:poisson}
and Section~\ref{sec:scattering}).  The investigation of the
mathematical and physical structures resulting from this basic program
is the subject of the remainder of this series of papers.  We shall begin,
however,  under the guise of attacking the concrete problem of one
harmonic oscillator dissipating (transferring) 
energy to another oscillator, and our
objective is to describe and constructively illustrate the structures
and methods which figure in providing mathematically respectable
solutions to that problem.

On the space $\F$ of the Gel'fand triplet $\rhs$, essentially self
adjoint operators need not be also be symmetric, i.e., may be 
non-hermitean.  When working in these RHS's the
usual notions of hermiticity or anti-hermiticity are not controlling,
and we will
need a radically different notion of what is the proper form of
adjoint involution in order to provide dynamical (including unitary)
representations~\cite{note3} of the
connected Lie groups which our complex Lie algebras--and associated
representations--integrate into through using the exponential map.  

Item 1 in the list above gives us
a more general set of solutions to work with.  Item 2, we will see,
gives us a geometric context in which it is mathematically proper to
speak of the Hamiltonian as the generator of infinitesimal time
translations (Item 5).  This interpretation of the Hamiltonian
does not automatically follow just because the
Hamiltonian has this meaning when restricted to another space (and
another topology).  This is also significant to establishing the
existence of Lie algebra valued connections, leading to a gauge
theoretic interpretation of some associated structures which we will
explore in the fourth installments of this series~\cite{IV}.
The third item is significant as indicating a possible source of
obstructions which must be avoided by proceeding carefully, and
our treatment demonstrates the inability of the standard Hilbert 
space formalism to address issues in the same
mathematical generality we have available in our RHS construction.

\subsection{Symplectic transformations of oscillators}\label{sec:sptrans}

In the non-relativistic RHS formalism, one works in a subspace of the
Schwartz space (of 
functions of rapid decrease) for the function space realization of the
abstract space $\F$ of the RHS $\rhs$.  The Lie algebra and Lie group
of symplectic transformations used in this present work have been 
associated with problems in quantum optics.  One therefore infers 
the methodology of this paper is likely to find concrete
application with electromagnetic fields.  The group of 
symplectic transformations is in fact
the group of squeezing transformations
of quantum optics, and provide the source of the transfer matrices of classical
optics in the paraxial approximation.  Hence, the present work can be considered the study of
the squeezing transformations of {\em generalized} Gaussian wave 
packets~\cite{note7}!  Physically, we are dealing with families 
of generalized Gaussian minimum
uncertainty states.  (If a field theoretic interpretation is adopted
for the oscillators, a natural candidate for the representation of a
``particle'' is a stable member of these families of minimum uncertainty
states.)   Two free oscillators are ``squeezed'' into correlation (or
coherence) by the symplectic (=dynamical)
transformations, and further ``squeezing''
results in resonant decay of the coupled oscillator
system.

\subsection{Order of proceeding}\label{sec:proceeding}
	
The appendix contain some important calculations.
Appendix~\ref{sec:DHO} recapitulates the Feshbach-Tikochinsky $SU(1,1)$
dissipative oscillator system calculation~\cite{ft} in slightly changed notation
(and with changed physical content, in order to put things in
a form which is not otherwise remarkable).  

We begin with a bit of naive algebra which requires substantial
justification, which we also begin in this section.  This is
a variant of the Feshbach-Tikochinsky calculation in which we take the
transformations to be part of a realization of the canonical
(symplectic) transformations for the two oscillator phase space and
{\em also define the action of the appropriate Lie
group (representation) to have a symplectic action on our spaces of
states}.  The simple
computation in Section~\ref{sec:scattering} has implications at
many levels.  It is shown that the dissipative oscillator system
can be obtained from a realization of the (semi-)group of 
symplectic transformations
applied to the system of two free quantum oscillators.  Further
(complex) symplectic transformations yield vectors which decay
exponentially (without regeneration).  It is also
demonstrated that similar constructions can generate a representation
of $SU(2)$ in terms of creation and destruction operators, and further
complex symplectic extensions will yield a complex spectrum for it 
as well, indicating a
quite general process is involved.  

The present results ultimately must be compared to the analytic 
continuation used by Bohm to obtain his Gamow
vectors~\cite{bohm1,bohm2}.  The two methods  
both describe complex symplectic transformations, so the necessary and
sufficient ``very well  
behaved'' starting point for obtaining the Breit-Wigner resonance
poles to associate to Gamow vectors~\cite{gadella} for the dissipative
coherence of two oscillators problem is not the F-T 
system of coupled oscillators, but should probably be thought of as 
the system of two free oscillators.  

Section~\ref{sec:poisson} provides a summary
descriptions of aspects of well known structures which will play
important roles for us.  These establish a basic general mathematical
context in which we operate.
In Section~\ref{sec:poisson}, the subject is the multitude of
symplectic and Poisson structures associated with representation of
both real and complex
semi-simple Lie algebras in a RHS format.  The structure of the 
space $\F^\times$ of the RHS $\rhs$ is not completely known, but 
in the case
of a Lie algebra representation, the structure of
$\F^\times_{\g\pm}$ follows the structure of $\g^\times$ in many
important regards, and there is some elaboration on this.  

The fact that the momentum map for a
semi-simple Lie algebra $\g$ is an element of $\g^\times$ provides us
with the first indication of a mandatory topology change to a
weak-$^\times$ topology: viewing a formal ``adjoint'' complex symplectic 
transformation on $\g= \mathfrak{sp}(4,\R )$, taking ``$Ad_G \g$'', as
transitive and also having a symplectic action on a space requires us
to view it as giving rise to a {\em co-adjoint} representation of
$\g^\C$ in $(\g^\C)^\times$.  This co-adjoint orbit structure within
the individual symplectic sheaves making up the Poisson manifold
$(\g^\C)^\times$ is 
faithfully reflected by the associated orbits of transformations on the
representation spaces $\F_{\g\pm}$ and $\F^\times_{\g^\C\pm}$.  By
appropriate identifications, the symplectic (and therefore Poisson)
mapping of 
inclusion (of the representation of $\g$) into these co-adjoint orbit 
structures (lying inside the representation of $(\g^\C )^\times$) 
can be made to provide essentially self adjoint extensions of
generators {\em and} a representation of the associated
transformation (semi-)groups whereby the exponential mapping for $\g^\C$ 
can be identified properly with the exponential mapping of 
$\g^\C \subset (\g^\C )^\times$. 

The representations of the associated simple
complex semi-groups are defined below in such a way that the
transformations have a symplectic action on the representation spaces
themselves.  (See Section~\ref{sec:scattering}).

The overall procedures closely follow the prescription in~\cite{reed}.  
There is a topology change mandated by {\em{continuous}} 
transformations of analytic continuation, 
and this point is very subtle and easy to overlook, but after 
the continuous analytic continuation one's solutions have been
extended to the distributions and only 
a weak-dual topology is appropriate~\cite{gadella}.  (Once again,
we're not in Hilbert space anymore!)

\section{``Scattering'' of Simple Oscillators}\label{sec:scattering}

In this section, we study the algebraic structure of the ``dissipative
oscillator'' system, but this section may also be interpreted as
a brief exercise undertaken to indicate the possibility of 
an algebraic theory of scattering, which is rigorous.  Such systems for the 
description of scattering have been considered before, e.g.,~\cite{gursey} uses
$SU(1,1)$ as the continuation of $SU(2)$, and uses $Sp(4,\R )$, to construct a 
partial theory of this sort.
Herein, a pair of free oscillators is subjected 
to canonical transformations to become an 
interacting (correlated or coupled) system in which one oscillator 
is ready to transfer 
energy to the other.  Further canonical transformations lead to
a dissipative oscillator system in which the flow of energy from one
oscillator (with higher energy) in the coupled system to the other
oscillator (with lower energy) is described by Gamow vectors which are 
energy eigenvectors with complex energy (and which therefore exhibit 
exponential growth or decay in their time evolution).  This is the result of 
the identification of the interaction Hamiltonian of the coupled system with 
a non-compact generator of the algebra $\mathfrak{su}(1,1) \subset
\mathfrak{sp} (4,\R )$, which is extended 
(analytically continued) from that real algebra to a complex algebra with the 
same generators and commutation relations defined (i.e., extended to a complex 
covering algebra).  This use of symplectic transformations is a
generalization of the F-T results recapitulated in
Appendix~\ref{sec:DHO}.  Any required additional justifications of the 
naive algebraic manipulations of 
the present section will be given in later sections.

The commutation relations useful to us 
for a unitary representation of $Sp(4,\RR )$ 
on Hilbert space, $\HH$, are~\cite{dirac} ($i,\, j,\, k=1,\, 2,\, 3$, sums implied):
\begin{eqnarray}
\left[J_i ,J_j\right] &=& \epsilon_{ijk} J_k   \label{eq:su2}   \\
\left[J_i,J_0\right] &=& 0                      \label{eq:zero} \\
\left[K_i,K_j\right] &=& -\epsilon_{ijk}J_k \\
\left[K_i,J_j\right] &=& \epsilon_{ijk}K_k \\
\left[Q_i,Q_j\right] &=& -\epsilon_{ijk}J_k \\
\left[Q_i,J_j\right] &=& \epsilon_{ijk}Q_k \\
\left[K_i,Q_j\right] &=& \delta_{ij}\, J_0 \\
\left[K_i,J_0\right] &=& Q_i \\
\left[Q_i,J_0\right] &=& -K_i	       \label{eq:lastcom}
\end{eqnarray}

An appropriate realization of this algebra in terms of two mode creation and 
annihilation operators is~\cite{dirac,kimandnoz}:
\begin{eqnarray}
iJ_1&=&{\frac{1}{2}}\;{\Big (}A^\dagger B + B^\dagger A{\Big )}   
							\label{eq:gen1} \\
iJ_2&=&-{\frac{i} {2}}\;{\Big (}A^\dagger B - B^\dagger A{\Big )}   
							\label{eq:gen2} \\
iJ_3&=&{\frac{1}{2}}\;{\Big (}A^\dagger A - B^\dagger B{\Big )}	  
							\label{eq:gen3} \\
iJ_0&=&{\frac{1}{2}}\;{\Big (}A^\dagger A + B B^\dagger {\Big )}   
							\label{eq:jzero} \\
iK_1&=&-{\frac{1} {4}}\;{\Big (}A^\dagger A^\dagger +  AA - B^\dagger       
	B^\dagger - BB   {\Big )}	\\
iK_2&=&{\frac{i}{4}}\;{\Big (}A^\dagger A^\dagger- AA +B^\dagger       
	B^\dagger - BB {\Big )}		\\
iK_3&=&{\frac{1}{2}}\;{\Big (}A^\dagger B^\dagger  + AB{\Big )}	\\
iQ_1&=&{\frac{i}{4}}\;{\Big (}A^\dagger A^\dagger -AA - B^\dagger       
	B^\dagger + BB   {\Big )} 	\\
iQ_2&=&-{\frac{1}{4}}\;{\Big (}A^\dagger A^\dagger +AA + B^\dagger       
	B^\dagger + BB   {\Big )}	\\
iQ_3&=&{\frac{i}{2}}\;{\Big (}A^\dagger B^\dagger  -AB{\Big )}  
							\label{eq:gen10}  
\end{eqnarray}

Identifying $X=iJ_1$, $Y=iJ_2$, and $Z=J_3$, we obtain the realization
of the $SU(1,1)$ Lie algebra generators used in Appendix~\ref{sec:DHO}.

The Baker-Campbell-Hausdorf relation  
\begin{equation}
e^B\, Ae^{-B} \, = \, \sum^\infty_{n=0} \;\frac{1}{n!} 
\left[ B,\,   \left[ B,\ldots \left[ B, A\right] \ldots \right] \right]
\end{equation}
containing $n$ factors of $B$ in each term, can be applied to semi-simple Lie 
groups and algebras, e.g., to $B = \mu X\in \mathfrak{su}(1,1) \subset
\mathfrak{sp} (4,\R )$, 
$A=Y \in \mathfrak{su}(1,1)\subset \mathfrak{sp}(4,\R )$, $\mu\in\R$, 
to yield:
\begin{equation}
e^{i\mu X}\left( iY \right) e^{-i\mu X}=\left( iY \right) \cos {\mu} - [X,Y] 
\sin {\mu}.
\end{equation}

For $\mathfrak{su}(1,1)$ and for pure imaginary $\mu$, the cosine becomes 
hyperbolic cosine (cosh) and sine becomes hyperbolic sine (sinh).  For the 
semi-simple group $SU(1,1)$ and its semi-simple algebra 
$\mathfrak{su}(1,1)$, it
 follows that:
\begin{equation}
e^{i\mu X}(iY) e^{-i\mu X}=(iY)\cos \mu -Z \sin \mu  \label{eq:adjoint}
\end{equation}
so that
\begin{equation}
e^{i({\pi}/{2}) X}(iY)e^{-i({\pi}/{2}) X}=-Z 
\end{equation}
or
\begin{equation}
Y=i\;e^{-i({\pi}/{2}) X}\, Z\, e^{i({\pi}/{2}) X}
\qquad Z=-i\;e^{i({\pi}/{2}) X}\,Y\,e^{-i({\pi}/{2}) X}   
\label{eq:F-T}
\end{equation}
(Note that we are not in Hilbert space, so that ``hermiticity'' is not
a relevant concept for us--it is not the case that 
conjugation of an hermitean operator by a seemingly
``unitary'' transformation has resulted in a non-hermitean operator.)  

Note the ``brazen'' use of the exponential map in the preceding. Full justification
for this will have to wait until installment four~\cite{IV}, but we will make some
preliminary comments here and further comment later on. The exponential map
is to be understood geometrically here, translating a tangent vector back and
forth along a geodesic (in the tangent space to the relevant group).  For now,
we will merely comment that this does not mean that we are defining a general 
inverse for any group element, but are working locally for limited displacements
along single geodesics only. In general, inverses may not be unique. However, 
the maximal sub-semigroup of the dynamical transformations - the semigroup of 
unitary transformations - differs from the unitary group by no more than a set of 
measure zero, so at least some of the dynamical transformations can be viewed as
fully invertible. In fact, we will see in part four of this series~\cite{IV} that conjugation
can be identified with a spin transformation acting from the left only, but preliminary to that
we must establish that the multicomponent vectors we define in this part are in fact
truly spinors, deal with the non-trivial issue of convergence of the exponential 
map. Physically, 
this approach is equivalent to insisting upon relative physical isolation: as time 
evolves, there are no additional dynamical interactions which further
change our Hamiltonian and
redefine the tangent to the geodesic of dynamical time evolution. The 
period of time of relative dynamical 
isolation may be billions of years for some photons to an astronomer or the
minutest fraction of a second in a high energy experiment. As a practical matter, 
some sort of relative causal isolation is required for any precision experiment, so,
although the mathematical meaning here may be a little vague (awaiting fuller
development in part four~\cite{IV}), the physical meaning is untroubling.

Note that for $\mathfrak{su}(2)$, where 
$\left[ J_k,J_l \right] = \varepsilon_{klm} J_m$, for real $\mu$
\begin{equation}
e^{i\mu J_k} \left( i J_l \right) e^{-i\mu J_k} = \left( i J_l \right) 
\cosh \mu -  \left[ J_k,J_l\right] \sinh \mu \quad .
\label{eq:su2adjoint}
\end{equation}
For pure imaginary $\mu, $ the corresponding expression  is 
$\left( i J_l \right) \cos \mu +i \left[J_k,J_l\right] \sin\mu$.  This is 
consistent with the $\mathfrak{su} (1,1)$ results, because there is a (
``dangerous to use'') mapping $\mathfrak{su}(2) \longrightarrow 
\mathfrak{su}(1,1)$ 
given by $J_1 \longmapsto X= iJ_1$, $J_2 \longmapsto Y=iJ_2$, $J_3\longmapsto 
Z$.  The Baker-Campbell-Hausdorf relation applies to both compact and 
non-compact groups, and to pure real and pure imaginary coefficients.  

This transformation procedure (which is based on the 
Baker-Campbell-Hausdorf relations) results in ``complex spectra for $SU(2)$'' 
just as a similar transformation resulted in ``complex spectra for 
$SU(1,1)$''. 
This illustrates that a general process is going on applicable to all locally 
compact semisimple Lie subgroups of a complex semisimple Lie covering group, 
and, in particular, applicable to both non-compact and compact 
sub(semi)groups alike.

Consider a system of two independent simple  
harmonic oscillators.  The Hamiltonian for this system is:
\begin{equation}
H = \frac{1}{2}\left( p^2_x + q^2_x +p^2_y +q^2_y \right).
\label{eq:Haf}
\end{equation}
This is the same operator as $iJ_0$, equation (\ref{eq:jzero}): $H=iJ_0$. 
If we subject this system to as ``preparation procedure'' the
symplectic (=dynamical) transformation:
\bea
&\, \alpha e^{i(\pi /2)J_1}&e^{i(\pi /2)K_2}\;(iJ_0)\; e^{-i(\pi /2)K_2}       
            \,\alpha e^{-i(\pi /2)J_1} \qquad 
	\nonumber	\\        
&\;&\qquad\qquad \;+\; \beta e^{i(\pi /2)Q_1}    \,e^{i(\pi /2)K_2}            
 \; (iJ_0)\; e^{-i(\pi /2)K_2}\, \beta e^{-i(\pi /2)Q1} = \quad                
         \nonumber   	\\   
\qquad &=&- \alpha e^{i(\pi /2)J_1} \, [K_2,J_0] \,                
		\alpha e^{-i(\pi /2)J_1} -\,         \beta e^{i(\pi /2)Q_1}  
		\, [K_2,J_0] \, \beta e^{-i(\pi /2)Q_1}               
	\nonumber   \\   
\qquad &=& +i\alpha e^{i(\pi /2)J_1} \; \left( iQ_2 \right) 		
		\;\alpha e^{-i(\pi /2)J_1}             
	\; +\; i \beta e^{i(\pi /2)Q1}\; \left( iQ_2 \right)
	\; \beta e^{-i(\pi /2)Q1}                 
		\nonumber  	\\    
\qquad&=& i\alpha^2 \; [J_1, Q_2]\;\; + \;\;              
	i\beta^2 \; [Q_1,Q_2] \nonumber  \\    
\qquad&=& i \alpha^2 \; (-\epsilon_{123}Q_3 ) \; +\; 
	i \beta^2 \; (- \epsilon_{123} J_3)                  
		\nonumber 	\\   
\qquad &=& -\alpha^2 \; \left( iQ_3 \right)  \; - \; \beta^2 \; 					\left( iJ_3 \right)         
\label{eq:adext} 
\eea
Choosing
\begin{equation}
\alpha^2 = -\frac{\G}{2} \qquad\qquad \beta^2 = -\Omega
\end{equation}
we have recovered the Hamiltonian of the $SU(1,1)$ dissipative oscillator 
system, equations (\ref{eq:ham1}), (\ref{eq:ham2}), (\ref{eq:ham3})  (See 
Appendix \ref{sec:DHO}).  Our Hamiltonian eigenstates have changed from 
$\{|n_A\rangle\}\oplus \{|n_B\rangle\}\in \F\cap\HH$ for the two free 
oscillators (energy/number representation) into two-component
state vectors representing mixed states which provide the foundations 
for the representations of the two  semi-groups $SU(1,1)_\pm\subset
Sp(4,\R)_\pm $. The function space realization of these extensions runs 
the from the space spanned by the direct sum of the 
``very well behaved'' free oscillator energy eigenstates representing 
the creation and destruction operator algebra, to the relevant
representations of $SU(1,1)^\C_\pm \subset
Sp(4,\R)^\C_\pm $.

Physically, the free oscillator system has been transformed by the
introduction of correlation into a system now composed of two
correlated oscillators--and the correlated system representation is 
split into a pair of semigroup representations split along time domains of
definition.  The states and related spaces are now mixed. 
That correlated system, in turn, is able
to represent further internal transformation representing the 
redistribution the energy
between the two oscillators.  See Appendix~\ref{sec:DHO}.  Note that
the correlated system acts as a (formally) closed system in which both source
and sink for the ``dissipation'' are dealt with. We have a system which
can not only represent the boundary conditions of an irreversible 
quantum dynamical process (without regeneration), but we can represent 
such a process in a dissipative context such as a thermodynamic reservoir
(because the annihilation operator is a bounded continuous operator in 
the appropriate topology, this reservoir can be infinitely deep).

In~\cite{lindblad}, use is made of rigged Hilbert spaces for
representation of $SU(1,1)$, and considerable care is taken to choose
a topology which insures any anomalous complex eigenvalues for this
real group are properly avoided.   
Herein, the anomalous complex eigenvalues are not excluded, meaning 
that at a minimum we must 
be working in some complex covering group, such as
$SL(2,\C )$ or $Sp(4,\R )^\C$, for which complex
scalars are defined--and we note that analytic continuation is a
complex symplectic transformation, which narrows our choices.  Now 
making an  
additional extension of the {\em full} Hamiltonian of the dissipative  
oscillator system to $\F^\times_{\mathfrak{sp}(4,\R )_\pm}$ using the 
$e^{i\mu K_3}$ map used in 
Appendix \ref{sec:DHO} (as an exemplar of the extension
process---there are nine other possible generators for the liftings
and other possible scalar coefficients), for the value $\mu =\pi /2$
we find: 
\bea
e^{i(\pi /2)K_3} \, H \, e^{-i(\pi /2)K_3} &=& \alpha^2           
	e^{i(\pi /2)K_3}\; \left( iQ_3 \right) \; e^{-i(\pi /2)K_3}
	     \nonumber   	\\       
& &\qquad +\, \beta^2 e^{i(\pi /2)K_3}\; \left( i J_3 \right) \;
	     e^{-i(\pi /2)K_3}                      
		\nonumber 	\\     
&=&- \alpha^2 \, [K_3, Q_3 ] \, - \, \beta^2 [K_3, J_3 ] 
		\nonumber 	\\     
&=& - \alpha^2 J_0 \; - \, \beta^2 (0) 
		\nonumber 	\\     
&=& i \alpha^2 \; \left( iJ_0 \right)   
\label{eq:Htransform}
\eea
$( iJ_0 )$ is the same as the realization of the
operator $iZ$ previously identified in the 
$SU(1,1)$ algebra, and this recovers the Feshbach-Tikochinsky result, 
equation (\ref{eq:F-T2}) or equation
(\ref{eq:F-T}).  There are other liftings, however, for which the commutator 
which results from the lifting of $H_0$ does not vanish.  
There are other decay 
constants (and decay processes) besides the single one identified in the F-T 
construction, although not all of these relate simply to the 
$\mathfrak{su}(1,1)$ algebra which the F-T procedure purports to be associated with.
(Reiterating, it is $\mathfrak{sp}(4,\R )^\C \cong \mathfrak{sp}(4,\C ) 
\supset\mathfrak{sl}(2,\C )\supset
\mathfrak{su}(1,1)$ which sets the overall structure.  This is only a
substructure of that larger $\mathfrak{sp}(4,\R )^\C$ structure.)~\cite{note16}

The essentially self adjoint extensions of the {\em generators} of the
realization of the Lie algebra can be understood as, for instance:
\bea
(iY)^\times = iY &=& e^{-i(\pi /2)K_3}\cdot\left[            
	e^{+i(\pi /2)K_3} \;(iY)\; e^{-i(\pi /2)K_3}  \right]          
	\cdot e^{+i(\pi /2)K_3}                      
		\nonumber  	\\     
&=& -\, e^{-i(\pi /2)K_3}Ze^{+i(\pi /2)K_3}
\eea
This amounts to identifying the simple inclusion of an operator
belonging to $\g_\pm$ into $(\g^\C_\pm )^\times$ with a certain point
on the co-adjoint orbit of another operator also within $(\g^\C_\pm
)^\times$, which can be explained as the failure of the
weak-$^\times$ topology to separate them.

Beyond those justifications for such an interpretation already given, 
we might add
that this is of the form $gAg^{-1}$, $g\in Sp(4,\R )^\C$, $A\in
\mathfrak{sp} (4,\R )$, and that on the semi-group representation space
for which $g$ is continuous, the operator $g^{-1}$ cannot 
be a continuous operator except for limited (finite) translations of the
relevant generators along the relevant geodesic, according to the exponential map.  
(The conjugacy of $iY$ and $Z$ will be
addressed at length in the fourth installment~\cite{IV} of these
articles, when we discuss the nature of the covering structure which
makes all the exponential maps well defined, and which 
in turn makes the structure spanned by the related
Gamow vectors with their associated Breit-Wigner resonances also well defined.) 
From the perspective of Lie algebra representations on
$\F\subset\F^\times$, esa operators need not
be symmetric, contrary to the situation in $\HH$.  As to $\F$ and
$\F^\times$, there is no violation of any sense of hermiticity in the 
above.  (Again, we're not in Hilbert space anymore, so insisting on
strictly applying notions of hermiticity or anti-hermiticity is unduly
restrictive.)

Many readers may be asking why we do not just go ahead and speak of
anti-self adjoint (anti-self dual, or asd) extensions.  The generators 
of the 
Lie algebra $\g$ are also canonically generators of the dual Lie
algebra $\g^\times$ due to the canonical $\pm$-inclusion
$\g\subset\g^\times$.  The generators of the algebra are taken
to be identified with generators of infinitesimal translations
(directional derivatives), and these have
invariant geometrical meaning as a certain tangent vector.  If we wish   
to identify representations of these generators 
in the Lie algebra with connections
associated to covariant derivatives, i.e., identify $U(1)$
subgroups as geodesic subgroups, then we must understand that 
the semi-simple Lie
algebras we are working with here may be associated with a group or 
with either of a pair of semi-groups, and the related duals: we are 
speaking of the same
generator in all cases.  (Because generators are first order 
derivatives, on analytic spaces the generator is locally 
independent of the direction of travel along the path of $exp$.)
Hence, because $\g \subset \g^\times$, the generators of $\g$ are
canonically (essentially) self adjoint (or self dual).  To
define a dynamical inner product transformation structure, on the other
hand, requires an anti-self adjoint (e.g., anti-self dual or asd)
extension of the Lie algebra~\cite{note17}.

It is necessary of symplectic transformations that their group obey
the dynamical law.  I.e., to say that a transformation $t$ belongs to
the group of symplectic transformations of a complex symplectic
space necessarily implies that $t\cdot t^\times$ must be the identity 
transformation~\cite{porteous}. This is a necessary and sufficient requirement.
Therefore, there is really no other alternative for the form of
adjoint involution other than the alternative chosen here, assuming we
persist in the requirement that our Hamiltonian generate infinitesimal
dynamical (=symplectic) transformations. We continue to mean that
the "inverse" here is to be understood only locally, for a period of
relative dynamical isolation.

For unitary transformations on Hilbert
space, one conventionally thinks of the dual transformation in terms
of hermitean (i.e., complex) conjugation, e.g., for esa Hamiltonian
\begin{equation}
\left( e^{-iHt}\right)^\dagger = e^{(-i)^\ast H^\dagger t} =
	e^{+iH^\dagger t } = e^{+iHt}
	= \textrm{``}\left( e^{-iHt}\right)^{-1}\textrm{''} \quad .
\label{eq:trad}
\end{equation}
The unitary transformations on $\HH$ are thus dynamical in their
action on $\HH$, and indeed the symplectic group contains a unitary
group as maximal compact subgroup.  This means the unitary
transformations form only
a proper subgroup of the group of all possible dynamical (symplectic)
transformations on $\HH$.  
On $\F$ and $\F^\times$, there one must think in terms of, e.g.,
\begin{equation}
\left( e^{-iHt}\right)^\times = \left( e^{- (iH) t}\right)^\times =
	e^{-(iH)^\times (-t)} = e^{+(iH)t} = e^{+iHt}
	= \textrm{``}\left( e^{-iHt}\right)^{-1}\textrm{''}
\label{eq:alttrad}
\end{equation}
because it must then also be the case that
\begin{equation}
\left( e^{Ht} \right)^\times = e^{H^\times (-t)} = e^{-Ht}
	= \textrm{``}\left( e^{Ht}\right)^{-1}\textrm{''}
\label{eq:realmod}
\end{equation}
in order that the group scalar product be properly
defined~\cite{note18,note18x5}.  
(The quotations on the right hand sides
above are cautions that care must be exercised interpreting these
relations, because, e.g., defining the rotation group by $RR^\dagger =
\II$ or by $RR^{-1}=\II$ leads to different spin groups covering the
rotation group.)  This should look more natural when we look at the
spinorial nature of these constructions in the fourth
installment of this series of papers~\cite{IV}. There are also 
topological obstructions to general inverses in the present setting,
as mentioned earlier. The back and forth along a geodesic must
be understood as mathematically local and relatively
isolated physically. In installment four~\cite{IV}, we will see that
our semigroup is made up of a series of local patches - one solves 
hyperbolic differential equation problems in such local patches (and 
worries about the permissible size of the patches a great deal) - and the
understanding we assert here amounts to working only within a
single of those local coordinate patches when we are describing
our dynamics in general, but that there is the possibility of a broader
description with regard to the areas of compact dynamics, e.g., 
resonances are locally restricted but you may work ``globally'' when
within the islands of stability providing you use the appropriate topology.

Recapitulating, in order to accomplish this properly dynamical
construction, we see from the above that must:
\begin{itemize}
\item{Deal with the complex Lie algebra as a {\em{real}} algebra.
Note that there is a unique decomposition of complex 
operators (e.g., a Lie algebra as an operator algebra) into their real or symplectic form,
 e.g., the complex Lia algebra $\mathfrak{g}_\C$ can be identified  with $\mathfrak{g}_\C  
\longrightarrow  \mathfrak{g}^\C \cong\mathfrak{g}\oplus i\circ\mathfrak{g}$~\cite{teichmuller,handb,adler}.  
Thus, we distinguish $H$ and $(iH)$ as
distinct generators of a real operator algebra containing $H$ itself
as a generator, because there is a complex plane structure present.}
\item{Accompanying the dual extension of this real operator algebra
must be an involution, transforming the real algebra used as 
semi-group ring scalars in such a way
that the resulting semi-group transformations satisfy the tests for
symplectic action.}
\item{These structures must be represented in representation spaces.}
\end{itemize}

This form of extension which is self-dual (esa) as to the generators
of the realization of the semi-algebra but anti-self-dual as to the 
realization of elements of the Lie algebra
and associated semi-group
sets up the connections due to the ``$U(1)^\C_\pm$'' sub-semi-groups 
which figure in the identification
of a generalized Yang-Mills gauge structure associated to the
group representations in generalized spaces used in 
this present body of work (see installment four~\cite{IV}
of this series),
and also figures in setting up the quasi-invariant measures which are
part of establishing the ergodicity of the Gamow vectors.  (See
installment three of this series~\cite{III}.) 

This esa extension makes possible the geometric identification of the
roles of the Lie algebra
generators in their role as generators of infinitesimal
translations in the related semi-group.  Their being esa also
enables us to apply the nuclear spectral theorem to the generators of
the Lie algebra, meaning when we proceed in the manner chosen for
the adjoint involution, the generators
can be associated with physical observables.  Thereby, we are
simultaneously 
creating an identification between dynamical quantities and physical
observables by means of careful mathematical construction.

The existence of a complex
structure on $\F_{\mathfrak{sp}(4,\R )^\C_\pm}$ (induced by the 
representation of complex Lie algebras 
$\mathfrak{sp}(4,\R )^\C\pm$) means,
for instance, that $Y$ and $iY$ define {\em{different}} connections
associated to different directional derivatives along transverse geodesics
on $\F_{\mathfrak{sp}(4,\R )^\C \pm}$, and each must be 
given an esa extension
individually.  It is obviously not possible to accommodate this structure
in the traditional Hilbert space formalism.  

There are {\em multiple} one parameter families
of symplectic esa extensions of the generators
from $\F$ to $\F^\times$, since there are 
other generators of symplectic transformations and the lifting 
parameters for all are continuous.  This is a 
radically different structure from  
the structure one is more familiar with in Hilbert space.  These group
extensions satisfy the often encountered and standard criteria for
``unitary representations'', e.g., if $\theta : G \longrightarrow
GL(V)$, then as a standard criterion for a unitary representation of 
$G$ on $V$ we require
$\langle \, \theta\circ g \, v \, , \, \theta\circ g \, w \, \rangle =
\langle \, v,w\, \rangle$, $\forall g \in G,\; \forall\, v,w\in
V$.  See, e.g.,~\cite{grillet}.  However, in the present context we see them as
dynamical (=symplectic) in their action rather than as unitary, since
the theorem of Porteous~\cite{porteous} applies to all the symplectic
transformations, of which the unitary transformations are a subpart.

For symmetric esa extensions (e.g., of hermitean operators 
on Hilbert space),
the eigenvalues in the unitary representation of the group are
necessarily unimodular, and the categorization of the present extensions as
quasi-invariant rather than invariant is intended to clearly signal that  
the eigenvalues of the group
representation need not be unimodular.  The scalar 
product is only ``quasi-invariant'' (see~\cite{genfun4}, page 311), even though
a test similar to the familiar test for unitarity is 
satisfied.  Following  the complexification--dual-extension, the
physical expectation values may be 
different from those previously obtained, since in general the 
extended operators
are no longer symmetric.  Eigenvalues for the quasi-invariant
semi-group
representation are not necessarily unimodular any longer, so that the
semi-group operator $e^{-iH^\times t}$ responsible for time evolution during
$t\ge 0$ need no
longer contribute a simple unimodular phase oscillation
for energy eigenvectors (Recall the Gadella diagrams in installment
one~\cite{I}--it 
is $H^\times$ and not $H$ which governs the time evolution during $t
\ge 0$) .

Many details will not be covered exhaustively herein, such as the 
existence of different left and right quasi-invariant measures, which
will be touched on very lightly
in the third and fourth installment of this series~\cite{III,IV}. The
``inverses'' associated with the quasi-invariant measures, e.g., the scalar 
product in our RHS representation, must again be understood locally rather 
than globally (if global, we would call them invariant). We are thus restricted
in how far our time parameter can run, for instance.

\subsection{Physical interpretation.}\label{sec:interp}

What is physically important are the matrix elements emerging from
this structure.  For a connected complex semi-simple Lie group $G$,
such as $Sp(4,\C )$, integration of forms on the 
group is equivalent to integration on forms on the algebra.  From bi-duality, 
one associates the scalar product $\langle g,h \rangle$ on $G_\pm $ and 
$G^\times_\pm$ with a scalar product on the representation space $V$, 
$\langle \bullet , \bullet \rangle_V$, $\theta : G_\pm 
\longrightarrow Aut(V)$, $\theta : \g_\pm \longrightarrow\mathfrak{aut} (V)$, 
by~\cite{warner}:
\begin{equation}
\left\langle \theta (\tau ) v ,\theta (\tau ) w \right\rangle_V= 
	\left\langle v,w \right\rangle_V \; ,\, \forall w\in V^\times \; , 
		\forall v\in V \; , \forall \tau \in G_\pm \; .
\label{eq:bidualityrep}
\end{equation}

For $G$ complex, there is a canonical scalar product on $V$ and 
$V^\times$ which is ``unitary'' (quasi-isometric):
\begin{equation}
\left\langle v,w \right\rangle_V = \int_{G\pm} \,
\left\langle \theta (\tau ) v ,\theta (\tau ) w \right\rangle_V \; 
d \tau
\label{eq:unitarygprod}
\end{equation}
This is a trivial extension of~\cite{warner}, p 151-153, to more general function 
spaces and to semi-groups.  Note that for the typical $e^{\pm iHt}$ time 
evolution semi-groups, this scalar product would be
phrased
\begin{align}
\int^\infty_0 \left( e^{-iHt} v\right)^\times \;\left( e^{-iHt} w \right) 
	d \mu_t    
& =\int^\infty_0 \left[ \left( e^{-iHt} \right)^\times v^\times \right]
	\;\left( e^{-iHt} w \right) 
	d \mu_t  				\nonumber 	\\
&= \left\langle v,w \right\rangle_V \;\; \int^\infty_0 \; d\mu_t 
	=\left\langle v,w \right\rangle_V
\label{eq:u1norm}
\end{align}
where $d\mu_t$ is the left invariant 1-form on the one parameter
time evolution sub-semi-group, properly oriented and normalized.  For a 
{\em{real}} semi-group generated by the essentially self adjoint operator $A$,  
one must make the association $\left( e^{\alpha A} v\right)^\times = 
 v^\times e^{-\alpha A}$, $\alpha \in\R_+$, in order for this to also result 
in a dynamical transformation of the inner product, i.e., physically one 
should think of the above integral as
\begin{equation}
\int_{G+} \, \left( e^{-iHt} v\right)^\times\, \left( e^{-iHt} w\right) \,
		 d\om_t 
		=\int_{G+}  v^\times e^{-iH^\times (-t)}\circ e^{-iHt} w\; 
			d \mu_t
\label{eq:ftnorm}
\end{equation}
or as the ``time reversed scalar product'' of F-T~\cite{ft} in some sense.

This mathematical structure (which is forced upon us in order to
provide both uniquely defined adjoints and
a proper dynamical representation of the connected complex
covering group)
has a sensible physical interpretation, as on our generalized space of
states it effectively measures 
the overlap of a state (=probability amplitude representing a dynamical
system) prepared during $t\le 0$ with the results 
of a measurement (=probability amplitude representation of a second
dynamical system used as a measurement apparatus) 
during $t\ge 0$.  Working in a rigged Hilbert space 
suggests it may be reasonable to adopt an untraditional 
but physically meaningful physical interpretation of evolution 
operators appearing in the integrals of evolving state vectors.  
This is an unexpected ancillary result of these 
quasi-invariant measures.

See also~\cite{butzer,bgm,bmlg} for extensive 
discussions of these semi-groups of time evolution and their physical 
interpretation.  This present work finally establishes they truly
deserve to be called semi-groups of dynamical
time evolution, and that they are not merely
generalizations gotten from the $U(1)$ groups of time evolution
obtained with the Schr{\"o}dinger equation in the HS formulation,
and for which the same physical meaning is conjectured.  Note that the 
{\em form} for the semi-groups of time evolution and their duals
obtained by formal analysis in~\cite{butzer,bgm,bmlg} is here obtained  
by an alternative means 
based on careful algebra and geometry.  

In an {\em algebraic} theory of scattering based on the RHS and this 
procedure, the M{\o}ller operators $\Om^\pm$ would be represented by
sums  and products of canonical transformations representing the
actual dynamical transformation processes, such as in
(\ref{eq:adext}).  (This represents a generalization of the {\em
  unitary} M{\o}ller operators.) The procedures dealt with here 
certainly do not constitute a complete theory of scattering, 
but do suggest that the formulation of quantum mechanics may be 
enlarged from a Hilbert space to 
a RHS to provide the vehicle from which a rigorous algebraic theory of   
scattering may be developed. 

The present algebraic approach to scattering is distinguished
by:
\bee
\item{Representation of the preparation and registration processes as  
continuous quantum dynamical transformations (which preserve some 
internal symmetry of the system).  There 
is no classical apparatus anywhere.}
\item{Dynamics can be separated into internal and external components, 
and both are based on symmetries.  The external evolutionary 
impetus takes the form of symplectic transformations which preserve 
some internal symmetry.  Internally, the identity of the oscillators
remains fixed by the internal symmetry (creation and destruction
operator algebra) which remains unbroken.  
The interpretation of this seems consistent with ``noisy'' external 
perturbations of the internal system and deliberate action upon the
internal  
system both being represented by canonical transformations. }
\item{There are no infinite reservoirs in the present construction,
  which involves a (formally) closed system of two oscillators
  dynamically evolving, and there is no privileged role for any
  apparatus or observer.} 
\item{The preparation or registration process is treated as a true
  dynamical process, driven in a self consistent way by the
  interaction of the system, and evolving continuously according to
  the appropriate $U(1)$ sub-semigroup of dynamical evolution.  Thus,
  for the preparation procedure represented by $e^{i\alpha A}$, the
  preparation advances with the transformation parameter $\alpha$.}
\item{Only essentially self adjoint operators are used as
  infinitesimal generators, and although the self adjoint Hamiltonian
  is transformed by the preparation procedure (and so may depend
  parametrically on the preparation process), it has no explicit time
  dependence.} 
\een
There are substantial open questions concerning the formalism, particularly 
relating to interpretation of the physical meaning of the mathematics, and, 
conversely, how to phrase a given physical situation mathematically.   
For instance, systems seem to evolve 
independently of any preparation or registration apparatuses, so one must 
distinguish the decay of a resonance from its observation, yet both seem to 
be representable by the same dynamical evolution parameter and 
transformation.  Apparently, one must encounter and properly interpret 
both active and passive canonical transformations, e.g., the decay
process itself might be considered passive and the registration of the
decay event by an experimental apparatus might be considered active.
A fuller account of stability against small perturbations also needs
to be provided for these quantum states which decay, and this will 
begin in the third installment of this series~\cite{III}.

\section{Poisson and Symplectic Structures}\label{sec:poisson}

The choice of spaces upon which to realize an 
operator representation 
of the algebras and semi-groups is very significant.  By a theorem of 
Marsden, Darboux's theorem does not hold for weak symplectic
structures~\cite{marsden}.
Thus, during the course of representing a Lie algebra structure 
(which participates in defining a weak symplectic form on
$\g\times\g^\times$) on a subspace of 
the Schwartz space (and its 
dual), Darboux's theorem is of no use to construct a local euclidean grid 
which is complete, e.g., there exists no way to completely diagonalize
the matrix elements of that representation
space and its dual (which is based on Darboux's
theorem).  This has the physical implication that there is a generic
possibility of mixing occurring during transformations on a space with
a weak symplectic structure.  (Mixing is a condition precedent for
entropy increase, discussed in installment three~\cite{III}.)

Semi-simple operators locally may be diagonalizable
individually (since they are ``linear''), 
but in the present situation 
there is no possibility of a complete set of commuting 
operators being obtained using Darboux's theorem.  (A complete 
set of commuting operators may still be obtained at the even higher
level of the universal enveloping algebra in the case of 
semi-simple groups.)   This weak symplectic structure
also permits the faithful representation of some 
sub-semi-groups which may contain elements not describable by the
exponential  
mappings of single generators~\cite{hilgert}, meaning we may add
certain sets (e.g., countable sets which are therefore of
measure zero) to our spaces without adverse effect on our mathematics.   

Marsden's Theorem means that there are non-trivial
local invariants (such as curvature or torsion) for 
the structure on $\F\times\FT$.  Thus, the antilinear
functionals $F(\phi ) = \langle \phi \vert F\rangle$ have just such
a symplectic structure, i.e., for $F=E$, the energy representation
in $\CS\cap \HH^2_\pm$ has a weak symplectic structure.  The existence 
of non-trivial local invariants
means the geodesics which represent the evolution of a state in the
space of states can be non-trivial geometrically, and the dynamical
physical evolution of the systems they idealize may be non-trivial as
well.

\subsection{Canonical Poisson and symplectic
  structures}\label{sec:canonical} 

This section is a brief recapitulation and reinterpretation
of standard results for Lie
algebras and their duals from the perspective of Lie-Poisson
structures.  See, e.g.,~\cite{mandr}.  It will serve as
an indication of the structure of $\F^\times_{\g\pm}$, which 
reflects the structure of $\g^\times$ in important
regards.  

There is a canonical symplectic structure on the direct product
of a Lie algebra with its dual, i.e., on $\g\times\g^\times$.  There
is a similar canonical symplectic structure on $\F_{\g\pm}\times
\F^\times_{\g\pm}$ and other pairings of locally convex 
vector spaces and duals, e.g., associated with the scalar product 
viewed as a Cartesian pair.  These are categorized as {\em weak}
symplectic structures.  Note that the {\em real} Hilbert space $\HH$
is associated with
a weak symplectic structure, but the complex Hilbert space
$\HH^\C$ is associated with a {\em strong} symplectic form defined by the
Hermitean product on $\HH^C \times \HH^\C$, which coincides
with the symplectic form on $\HH^\C \times (\HH^\C )^\times$,
so a complex Hilbert space is associated with a strong symplectic form while a
real Hilbert space possesses a weak one.  (See, e.g.,~\cite{mandr}, chapter 2.)

The dual of a Lie algebra is a Poisson manifold, e.g., $\g^\times$ is
Poisson, and co-adjoint orbits in $\g^\times$ individually 
possess a symplectic
manifold structure: $\g^\times$ can be thought of as a union of these
co-adjoint-orbit-symplectic-manifolds (but need not be a symplectic
manifold itself, since those unions may be disjoint, meaning there is
no global symplectic form).  On $\g^\times$, we have a Poisson
structure associated with the Lie-Poisson brackets on $\g^\times$, and
a symplectic structure associated with the Lie-Poisson structure on
the co-adjoint orbits.  We thus have symplectic sheaves in the Poisson
manifold $\F^\times_{\g^\C\pm}$, e.g., ``$Ad_{Sp(4,\C )}$'' applied to the
realization of $\mathfrak{sp}(4,\R )^\C_\pm$ on
$\F_{\mathfrak{sp}(4,\R )^\C \pm}$
forms a series of symplectic sheaves (co-adjoint orbits) 
in $\F^\times_{\mathfrak{sp}(4,\C
)\pm}$.  Inclusion is a Poisson mapping.  We can also think of
$\g^\times$ as $T^\times G/G$, where $G$ is obtained by integration of
$\g$, and $\g^\times $ must be even dimensional~\cite{mandr}, page 293 and
Chapter 13.  For $\g$ the value of the {\em momentum mapping}
is always an element in $\g^\times$, and under the momentum mapping a
Poisson (e.g., including symplectic)
action of a connected Lie group $G$ is taken as the co-adjoint
action of $G$ on the dual algebra $\g^\times$.  See~\cite{arnold}, Appendix 5,
page 374.  

We thus have an association of transitive action of a
connected Lie group,  the momentum map, Poisson action of a group on a
space, and a 
topology change to a weak-$^\times$ topology in the dual of the Lie 
algebra of that Lie group.  The transitive symplectic action (momentum
map) of a
connected Lie group is taken as the co-adjoint action of that Lie group
on the appropriate dual.  We will see this structure mirrored every
time when we address the issue of why what appear to be
adjoint orbits on some representation space 
$\F_{\mathfrak{sp}(4,\R ) \pm}$ are in fact
co-adjoint orbits in the representation space
$\F^\times_{\mathfrak{sp}(4,\R )^\C \pm}$.  It is appropriate to begin
addressing that issue now.  The scalar product 
$\langle\,\bullet\,\vert\,\bullet\,\rangle$
on $\F\times\F^\times$ is what matters here, both physically and
mathematically.  There is a canonical symplectic structure here
(e.g., when viewed as a Cartesian pair), 
and it exists canonically on $\g\times\g^\times$ as well (provided
those transformations have well defined adjoints).  This 
means that, e.g., even a seemingly 
``adjoint transformation on $\g$'' can, if transitive, 
be thought of as an element of the group of
symplectic transformations on $\g\times\g^\times$, and that group of
(complex) symplectic transformations may include analytic continuation 
transformations (a form of complex symplectic transformation).  For
instance, even 
the mere choice of a ladder operator basis for a real group may amount to a
complex symplectic transformation in precisely this sense.  See, e.g.,
the $SO(3)$ and $SO(2,1)$ comparison, yielding a complex spectrum for
$SO(2,1)$ in this regard, used as an example in~\cite{bgm}, Section
2.

Gadella has given the necessary and sufficient mathematical conditions
for analytic continuation in the function space realizations of the
RHS structure~\cite{gadella}, and in particular shows that the analytic
continuation proceeds from the realization of $\HH^\times$ to the
function space realization of $\F^\times$.  H{\"o}rmander~\cite{hormander}, gives
a prescription for the analytic continuation of an elliptic Green's
function which demonstrates how one must proceed with the abstract
spaces.  The fundamental solutions are analytically continued, but a
topology change is mandated since this analytic continuation is to the
distributions.  Hence, a complex symplectic transformation of a real
algebra which works an extension to the complexes must be
accompanied by a topology change to a weak-$^\times$ topology.
Analytic continuation is a momentum mapping (!) in the jargon of
dynamical systems.  Tied
up in this general topology change shown by H{\"o}rmander are issues
in harmonic analysis necessary for the existence of Green's functions,
resolvents (whose poles provide the spectrum), etc.  All
the mathematical machinery necessary for us to do physics depends 
on us making
this topology change, and this topology change is not optional, but is
necessary for well defined mathematics as well as dynamics as we 
have sought to describe them.

\subsection{Symplectic structures and forms of
  algebras}\label{sec:spstructures} 

A complex structure is a type of symplectic structure.  It is
important in the present context to think of the complex numbers in
their symplectic form, i.e., as a real algebra $\R \oplus
i\circ\R$.  This is because if we think of the {\em field} $\C$ as an
algebra, the operation of 
involution in the form of complex conjugation cannot be 
uniquely be defined without the addition of more structure than 
the axioms of a field provide.  A space is an algebra plus a scalar
product, and when there is an inclusion of the space in the dual, as
there is in the RHS paradigm, the adjoint operation of the algebra
must be an involution, which suggests we should think of the ring of
scalars as an algebra and the adjoint operation as accompanied by an
involution of the algebra of scalars.  (This suggestion becomes
compulsory when one also wishes to define transformations in such a
way that their action is symplectic (=dynamical).  See below.)  For
example, if one thinks 
only  in terms of complex conjugation of the field of complex numbers 
on a conventional Hilbert space, the
representation of a complex Lie group obtained from exponentiating
a complex Lie algebra (using an esa realization of the generators of
the algebra, so as to make them identifiable with physical observables)
would make the scalar product of that space
not be uniquely defined, since exponentials of complex operators 
(e.g., with mixed real and imaginary parts) would not have unique
adjoints!  Those operators of the form $H=H_0 +iH_1$
could no longer be characterized as either hermitean or
anti-hermitean, and so any effort to accommodate complex operators and
complex spectra on von Neumann's Hilbert space is destined for very
serious mathematical difficulties. 

The usual insistence on hermitean (or anti-hermitean)
representations does not stem
from any structure of the Lie algebra, but from the need for unitarity 
in commonly met examples involving the von Neumann Hilbert space, with
the hermitean scalar product.  The hermitean conjugation there has a
symplectic (=dynamical) action on the von Neumann Hilbert space.
The extension of this form of hermitean scalar product
to our complex group/algebra situation on our different
spaces would not make the transformations represented complex
symplectic, however, which may be even more serious than being
non-hermitean and non-unitary, since this
relates to integrability in a more general way, and would also be
contrary to our current understanding of what is ``dynamical''.    

If, however, one thinks
of $\C$ as a real algebra (i.e., considers it in its symplectic form
$\R\oplus i\circ\R$), then one may define an involution for that real
algebra uniquely.  There may be many possible alternative ways of 
defining involutions for a given real algebra, but you must choose 
one form of involution which is unique and stick to it~\cite{note19}.  
In the interests of proceeding
uniquely in the present algebraically oriented treatment, it is
critical to think of ``analytic continuation'' not in terms of 
``field extension'' but in terms of (transitive)
transformations extending a real algebra into the symplectic form of an 
enlarged algebra.  We further insisted on defining the action of
representations 
of groups of transformations on our spaces in such a way that their
action on our representation spaces is symplectic~\cite{note20}.  
(See  again Section~\ref{sec:scattering}.)

A complex manifold 
with a hermitean metric whose imaginary part is closed (e.g., symplectic) is 
called a K{\"a}hler manifold.  On the complexification
of the real phase space $T^\times \R^n\cong\R^{2n}_{p,q}$, which will be 
denoted (abusively)
$\widetilde{T^\times \R^n} \cong \widehat{\R^{2n}_{p,q}}\cong \C^n$, 
one has an hermitean metric:
\begin{equation}
h({\boldsymbol \xi},{\boldsymbol \eta}) = 
	\left({\boldsymbol\xi},{\boldsymbol \eta}\right) - i 
		{\boldsymbol \om}({\boldsymbol\xi},{\boldsymbol \eta}) =   
\left({\boldsymbol \xi},{\boldsymbol\eta} \right) 
	- i \left[ {\boldsymbol \xi},{\boldsymbol \eta} \right] \; .
\end{equation}  
That $\boldsymbol \om$ is closed on $\R^{2n}_{p,q}$, $d{\boldsymbol \om} =0$, 
is a consequence of the existence of a symplectic (Darboux) coordinate system, 
i.e., Euclidean structure on $\R^{2n}$.  In such a coordinate system the
torsion vanishes, meaning (pseudo-)Riemannian connections can be defined for 
the manifold which possesses a complex structure~\cite{nash}, page
167f.  We may take the vanishing of the torsion for 
as the integrability condition for integrating Lie algebras to the 
(connected part of the) Lie group, i.e., so that torsion free
connections can be said to exist on the complex simple Lie group.

Given that the work herein involves the use of hermitean metrics on
complex manifolds, i.e., involves (weak) K{\"a}hler
metrics, the vanishing of the torsion is not a trivial 
matter, and depends on the fact that $\R^{2n}$ is torsion free, and that the 
complex semi-simple Lie algebras are torsion free.  In a generalized
Riemannian geometry setting, the skew symmetric part of
the metric is associated with possible torsion (and some corrections 
to the curvature tensor arising from the torsion), and the hermitean 
and K{\"a}hler metrics have just such a skew symmetric part, as shown
above.  The torsion itself is not an especially  
serious obstruction to integrability (e.g., we are untroubled by
holonomy, and indeed recognize it in physics as the geometric, or 
Berry, phase), but the conditions necessary for 
the existence of torsion also makes 
possible {\em shear}, and the formal possibility of essential
discontinuities on a set of positive measure can present serious
problems!  Local integrability is possible because
the possibility of shear on sets of positive measure is avoided through
satisfaction of the integrability condition on the geodesic semigroup.  We may view
shear as being restricted to, at most, a set of zero measure, which we
understand to be reflected in the branch cut (if any) 
and countable set of resonance poles associated with the analytic
continuation.  The branch cut represents a bifurcation of solution sets  
by discontinuities in the second and higher derivatives, and so does not
represent shear in the ordinary sense.  Hence, shear occurring during
the dynamical time evolution in our spaces of generalized states, if
present at all, exists only on a set of measure zero, the countable
set of Gamow vectors, or their associated countable set of
Breit-Wigner resonance poles~\cite{note21}.

\subsection{Hamiltonian symplectic actions (integrability)}
	\label{sec:hamiltonian}

Because we have defined a new type of Lie algebra and Lie group
representation, it is appropriate to take an aside to demonstrate that
our constructions are non-pathological.  We have constrained ourselves
by insisting that the 
representation have a symplectic action on the representation spaces,
and also that there be weak symplectic forms, etc., so some further
explanation is necessary.

The representation of evolutionary flows generated
by semisimple symmetry transformations are defined over a variety of spaces
representing quantum mechanical states, e.g., all the spaces in the
Gadella diagrams of installment one~\cite{I}.  There are a variety of
integrability conditions which must fully mesh everywhere throughout this
structure.  The conditions for the Lie algebra structure of the
infinintesimal generators of these transformations to be integrable
(e.g., into the connected part of a group structure) are reflected in
their representation counterparts as conditions for the flows to have
``single valued hamiltonian functions''.

For a semisimple 
symmetry group $G$, by limiting consideration
consideration to semi-groups and sub-semi-groups of $G$
which are strictly infinitesimally generated, we require in the first
instance that the group 
is the union of the two strictly infinitesimally generated 
semi-groups, which are unique: $G= G_+ \cup G_-$~\cite{hilgert},
p. 378. The 
representation of these strictly infinitesimally generated 
sub-semi-groups will be represented by
the Lie algebra as a subscript, as in $\CS_{\g\pm}$.  The 
complete spaces $\CS_{\CG\pm}$ and $\CS^\times_{\CG\pm}$ are not
necessarily 
restricted by this limitation of consideration, so that we must think 
$\CS_{\g\pm}\subseteq \CS_{\CG\pm}$, since there may be sheet to sheet jumps
within $\CS_{\CG\pm}$.  We are interested in the
intersection of the Schwartz space and space of Hardy class functions,
$\CS_{\g\pm} \cap \HH^2_\pm$ and $\CS_{\CG\pm}\cap\HH^2_\pm$,  
and with the $\HH^p$ spaces one cannot be assured that 
sufficiently many functionals  
exist to separate a point from a subspace~\cite{duren}.  The spaces of
physical interest, $\CS_{\CG\pm}$ and $\CS_{\g\pm}$, are likely 
to differ at most on a set of measure zero.  

A Lie algebra $\g$ when viewed as a linear 
vector space possesses a dual $\g^\times$, and one normally defines     
an inner product $\langle g , \xi \rangle$, $\xi \in \g^\times$, 
$g\in \g$.  (A linear space is an algebra plus a scalar product.) On
representation spaces,  
one conventionally reflects the bi-duality of this product 
(reflexitivity of $\g$ and $\g^\times$ as spaces) by appropriate 
definition of product modules over the representation spaces, 
and there should also be reflexitivity of the representation spaces
and duals~\cite{note21x5}.  For semi-groups and their algebra of generators, a 
similar scalar product can be defined, although one works with
semi-group rings rather than rings.  See Section~\ref{sec:scattering} and 
installment three of this series.  See also, e.g.,~\cite{yabogr}, page
398.

The continuous extensions of infinitesimal generators  $A \longrightarrow 
A^\times$, which are taken to be essentially self adjoint, is adopted
because the operator 
$A^\times$ dual to a generator $A$ is equal to the weak-$^\times$ 
generator of the dual semi-group.

When the $G$-action of a semi-simple group $G$ 
preserves the symplectic form $\boldsymbol \om$, all of the
fundamental vector fields of the action of $\g$ on $G$ are locally 
hamiltonian (locally integrable)~\cite{souriau}, p. 47, see
also~\cite{moser,siegel}.  The action  
of the strictly infinitesimally generated semi-simple (sub-)semi-groups 
thus preserve the symplectic form and all of the fundamental 
vector fields are thereby locally hamiltonian.  This follows because
$\g\subset\g^\times$, fixing the symplectic form on all of the
symplectic sheaves of $\g^\times$ (even though there may be no global
symplectic form on $\g^\times$ itself.)   

The Lie algebra of generators of 
the strictly infinitesimally generated Lie semigroups (as ray
semi-groups~\cite{hilgert}), 
therefore exponentiate to generate locally hamiltonian 
semi-flows on the (connected part of the semi-simple) semi-groups $G_\pm$.  
This locally hamiltonian structure should be carried forward by 
non-pathological representation mappings.  There may even be global
integrability, but at least
local integrability is assured for all semi-simple semi-groups and
their non-pathological representations.

At this juncture, we see torsion free structures on the complex
semi-simple Lie algebras whose groups are therefore (locally path)
connected (although
not necessarily simply connected).  We have symplectic structures on
the direct product of 
(complex) spaces and their duals.  We are thus ready to invoke the
result that the symplectic actions of semi-simple Lie algebras on
symplectic manifolds are Hamiltonian~\cite{arnold}, Appendix 5,
~\cite{souriau}.  In fact, as to 
the symplectic transformations on $\g\times\g^\times$ (e.g., on
$\mathfrak{sp}(4,\C )\times\mathfrak{sp}(4,\C )^\times$) we have a
textbook Hamiltonian action~\cite{arnold}, page 346, example (e):
\begin{quote}	
The co-adjoint action of $G$ on a co-adjoint orbit in $\g^\times$
with the $\pm$ orbit symplectic structure has a momentum map which is
the $\pm$ inclusion map\ldots This momentum map is clearly
equivariant, thus providing an example of a globally Hamiltonian
action which is not an extended point transformation.  
\end{quote}
Equivariance of the mapping is a condition for integrability.
See also Section~\ref{sec:scattering} and installment three~\cite{III} 
of this series.  The Singular Froebenius Theorem permits 
integration in systems involving distributions.

This globally 
Hamiltonian action in the abstract Lie algebra dual is reflected
in the representations.  In particular, the Hamiltonian vector fields
on the (weak) symplectic representation manifold have symplectic
flows upon which ``energy'' is conserved~\cite{arnold}, page 566.  This will
enable us to, e.g., make statements of a thermodynamic character
appropriate to an energetically isolated system of quantum resonances
represented by Gamow states, as we will do in the third installment
of this series.  This also
makes conservation principles available generally.

This globally Hamiltonian action may also exhibit the 
sensitive dependence on initial conditions (inherent in analytic
continuation~\cite{mandf1}).  Thus, in the present instance
the global integrability which the
momentum map promises may not be usable as a practical matter, and
in the case of analytic continuation, only local
integrability may be obtained as a useful result.

\section{Predicted Observables}\label{sec:observables}

There are several general areas in which possible observable consequences of 
this model may emerge, and these will be pursued in detail separately from 
this paper.  The most obvious is the quantization of the width (or, 
equivalently, the decay rate) as characterizing the resonances described 
here.  Widths are usually difficult to measure, so the actual experimental 
resolution of this effect may not be easy.  Note, however, that the $Z$ 
eigenvalue appearing in the width is also a characteristic of the 
$|j,m\rangle$ eigenstates prior to analytic 
continuation.  The realization of the $SU(1,1)$ algebra used herein (and also 
the $SU(2)$ algebra) has an association with interferometry in quantum 
optics~\cite{yurke}, and perhaps sufficient sensitivity can be attained by the
use of non-linear optical phenomena, such as four wave mixing, etc.  
The indicated decay rate is inversely
proportional to the total energy in the system (including the 
$\frac{1}{2} \hbar \omega$ vacuum zero
point energy) and independent of the energy gap between the two
oscillators (or fields). 

Because the
energy is analytically continued to include values from the negative real
axis, the excitation numbers of the modes can be negative.  (Recall that
$m=n_a +n_B$ is a characteristic quantum number
of the eigenstates of $SU(1,1)$ prior to the
analytic continuation which produced the Gamow states.)  For vacuum states,
$n_A +n_B =m =0$, there is still a finite decay probability for the
process transferring energy from oscillator $A$ to oscillator $B$.  This
provides an explicit representation of, e.g., the well known vacuum
fluctuations of quantum optics.   

The present work also implies the possibility of observing quantization
effects in the rate of decay of any two field 
resonant decay processes.  If one interprets the two field
interferometry constructions of~\cite{yurke} quite literally, then at the
appropriate energy scales quite general combinations of fields 
should show analogues to the familiar
interferometry process in quantum optics.  Thus, one infers that there
may be electroweak analogues to Fabry-Perrot interferometers, four
wave mixing, etc., at the electroweak unification scale of energy.    

It is also predicted that the characterization of an irreversible process 
can take a particular form: pure exponential decay.  In the derivation of 
the complex spectral resolution, Bohm obtained
two terms~\cite{bohm1,bohm2}, rather than the single sum over Gamow vectors obtained here.
Thus, for the prepared state $\phi^+$, one has formally
\begin{equation}
\phi^+ = \sum^n_{i=1} \psi^G_i \langle \ts^G_i | \phi^+\rangle	+ 
\int^{-\infty_{II}}_0  dE \;\; |E^-\rangle	\langle^+E|\phi^+ \rangle  
\;\; .
\label{eq:spectresolution}
\end{equation}
The first term on the right hand side corresponds to the result of the
present paper.  The second term on the right hand side (sometimes called
the ``background term'') has the geometric meaning of an holonomy 
contribution: there is no observable holonomy in the present two oscillator 
system since the evolution does {\em not} generate a 
trajectory which closes in the curved state space.  This, of course,
is precisely the sort of complication we avoided by working locally and
proscribing general inverses in the earlier chapters. The 
plain geometric implication is that even if one should 
manage to return to what might be called the initial configuration, the various
symmetry operations representing the necessary preparation steps and 
the necessary evolution steps will be 
by geodesic transport over curved space and one expects that some geometric 
phase (holonomy) will be accumulated, making the ``background term'' no 
longer zero.  Indeed, there may possibly be some general identification
between simple irreversible dynamical time evolution and holonomy.

It is to be expected that careful examination of the 
``background term'' should disclose the role of the history of the system 
(non-Markovian dynamical time evolution--discussed in part
three~\cite{III}) in the irreversibility of the system  as expressed by
the semigroups of evolution of that system.  One is thus  
led to expect deviations from exponential decay only in systems which
have a history, e.g., correlations among events between subsystems.  The
formalism predicts, e.g., that ``spontaneous'' decay should be
exponential, perhaps as the result of some stochastic external perturbation 
(see~\cite{III}), and further one might observe 
deviations from the exponential when there exists correlations
(coherence) between the
decay events.  This agrees with physical intuitions, and 
the principles seem to have weak (non-quantitative)
experimental support.  See~\cite{note22}.

The gauge interpretation of the symplectic transformations of this
paper which will be undertaken in the fourth installment of this
series should also have physical observables~\cite{IV}.

Finally, in resonant quantum microsystems, geometric phase (holonomy)
accumulating during system evolution seems inextricably associated
with intrinsic microphysical irreversibility.  For resonant quantum
quantum microsystems, there is perhaps some intrinsic association
between the accumulation of geometric phase (holonomy) and entropy
growth.

\appendix
\section{The $SU(1,1)$ Dissipative Oscillator}\label{sec:DHO}

This section is a partial recapitulation of the $SU(1,1)$ dissipative
oscillator, first described in the paper of Feshbach and Tikochinsky~\cite{ft}.  In 
addition to changes in emphasis and mathematical detail, in the present work 
there are changes of form, e.g., only operators 
which are ``symmetric under time reversal'' will be used
(contrary to the original).  From these 
operators which are time reversal symmetric, irreversible time evolution will 
emerge in the form of exponentially decaying Gamow vectors.  

The equation of motion of the damped simple oscillator is:
\begin{equation}
m{\ddot{x}} + R \dot{x}+kx=0
\end{equation}
Canonical quantization requires a Lagrangian, and the appropriate Lagrangian 
requires an auxiliary variable $y$:
\begin{equation}
L=m\dot{x}\dot{y} +\frac{1}{2} R(x\dot{y}-\dot{x}y)-kxy
\end{equation} 
The equation of motion for $y$ becomes:
\begin{equation}
m\ddot{y}-R\dot{y}+ky=0
\end{equation}
The Hamiltonian $H$ is
\begin{equation}
H=\frac{1}{ m} p_xp_y + \frac{R}{2m} (yp_y-xp_x) + 
	\bigl( k - \frac{R^2}{2m} \bigr) xy
\label{eq:Hint}
\end{equation}
The $[a,\,a^\dagger ]=\II =[b,\,b^\dagger ],\;[a,\, b]=0$, etc. commutation 
relations are equivalent to the $[p,\,q]=-i\hbar\,\II$, etc., Heisenberg 
relations, and canonical quantization is straightforward. After canonical 
quantization and translation of variables into destruction and creation 
operators for two modes, $a=(\sqrt{m} \Om x+ip_x/\sqrt{m})/\sqrt{2\Om}$,and 
$b=(\sqrt{m} \Om y+ip_y/\sqrt{m})/\sqrt{2\Om}$, rearrangement suggests the 
further change of variables:
\begin{equation}
a=\frac{1}{\sqrt{2}} (A+B)\;\qquad b = \frac{1}{\sqrt{2}} (A-B)
\end{equation}  

We then have a splitting of the Hamiltonian
\begin{equation}
H=H_0+H_1     
\label{eq:ham1}
\end{equation}
where
\begin{equation}
H_0= \Om (A^\dagger A-B^\dagger B) \qquad\qquad \Omega =k-\frac{R^2}{2m}    
 \label{eq:ham2}
\end{equation}
and
\begin{equation}
H_1=i\frac{\G}{ 2} (A^\dagger B^\dagger -AB)\qquad \qquad \G =\frac{R}{m}     
\label{eq:ham3}
\end{equation}
There is a natural system--reservoir interpretation of the above, and in the 
limit $R\rightarrow 0$ the eigenstates of $H$ reduce to those of the simple 
undamped harmonic oscillator provided one considers only states annihilated 
by $B$~\cite{ft}. 

The operator $H_0$ is simply related to the ${\mathscr{C}}^2$ Casimir 
of $SU(1,1)$.  The operator $H_1$ together with two other operators form a 
realization of the algebra of $SU(1,1)$, useful in determining the 
eigenvalues of the Hamiltonian:
\begin{equation}
iX=\frac{1}{2} (A^\dagger B^\dagger +AB) 
\label{eq:su11x}
\end{equation}
\begin{equation}
iY=\frac{i}{2}\; (A^\dagger B^\dagger -AB) 
\label{eq:su11y}
\end{equation}
\begin{equation}
iZ=\frac{1}{2} (A^\dagger A +BB^\dagger  )  
\label{eq:su11z}
\end{equation}
{obeying the algebra}
\begin{equation}
[X,Y]=Z
\label{eq:xyz}
\end{equation}
\begin{equation}
[Z,Y]=X
\label{eq:zyx}
\end{equation}
\begin{equation}
[X,Z]=Y
\label{eq:xzy}
\end{equation}

$iZ$ is essentially the Hamiltonian for the two mode simple oscillator, with 
eigenvalues $2m+{1}$, $m=({1/2}) (n_A +n_B)$, while we may label the 
eigenvalues of $H_0$ by $2\Om j$, where $j={1/2}\, (n_A-n_B)$.  

The Baker-Campbell-Hausdorf relation can 
be applied to $B = i\mu X\in SU(1,1)$, $A=Y \in SU(1,1)$, $\mu\in\R$, to yield:
\begin{equation}
e^{i\mu X}iYe^{-i\mu X}=iY\cos {\mu} - [X,Y] \sin {\mu}.
\end{equation}
For the semi-simple (e.g., non-solvable) group $SU(1,1)$and its semi-simple 
algebra $\mathfrak{su}(1,1)$, it follows that:
\begin{equation}
e^{i\mu X}iYe^{-i\mu X}=iY\cos \mu -Z \sin \mu
\label{eq:adjoint2}
\end{equation}
so that
\begin{equation}
e^{i({\pi}/{2}) X}iYe^{-i({\pi}/{2}) X}=-Z 
\end{equation}
or
\begin{equation}
Y=i\;e^{-i({\pi}/{2}) X}\, Z\, e^{i({\pi}/{2}) X}\qquad 
	Z=-i\;e^{i({\pi}/{2}) X}\,Y\,e^{-i({\pi}/{2}) X}   
\label{eq:F-T2}
\end{equation} 

Because the non-unitary dynamical transformations (analytic continuation)
work a complexification of the algebra 
$\mathfrak{su}(1,1) \longrightarrow \mathfrak{su}(1,1)^{\C}= 
\mathfrak{sl}(2,\C)$, the adjoint transformations of $(iY)$ exit 
the realization 
of $\mathfrak{su}(1,1)$ to become a realization of $\mathfrak{sl}(2,\C )$; 
the transformed eigenvectors have left the representation space for 
$\mathfrak{su}(1,1)_\pm$ into a representation space of 
$\mathfrak{su}(1,1)^{\C}_\pm= \mathfrak{sl}(2,\C )_\pm$ as well.  $H_0$ 
of the Hamiltonian is in fact proportional to the angular momentum operator 
$J^2$ of $\mathfrak{su}(2)$, and there is also 
the ``dangerous'' so-called analytic continuation of
$\mathfrak{su}(2)\longrightarrow \mathfrak{su}(1,1)$ given by 
$J_1\mapsto iJ_1=X$, 
$J_2\mapsto iJ_2=Y$, and $J_3 \mapsto Z$, so that $Z$ can also be thought of 
as an $\mathfrak{su}(2)$ generator, and thereby the source of a 
discrete spectrum; because $Z\in \mathfrak{sl}(2,\C )$, so is $iZ$, and 
hence the appropriateness of 
complex eigenvalues for $iZ$ on the eigenvector of $J_3$.  The complex 
eigenvalue is appropriate in $\mathfrak{sl}(2,\C )$ because it is a complex 
algebra.  

If the eigenstates of $Z$ are $|j,m\rangle\in \CS$, the Schwartz space, as 
above, the eigenstates of $Y$ resulting from the extension of these vectors 
to $\F^\times$ using the $Ad_{exp(\mu X)}$ map corresponding to 
$\mu=-{\pi}/{2}$ are $e^{-i({\pi}/{2})X}\,|j,m\rangle$:
\begin{equation}
\left( iY\right) \,e^{-i({\pi}/{2})X}\,|j,m\rangle=i\,(m+{1}/{2} )\,
	e^{-i({\pi}/{2})X}  \,|j,m\rangle 
\label{eq:la}
\end{equation}
Because $m\ge 0$, this is a positive pure imaginary eigenvalue.  There is 
also a negative pure imaginary eigenvalue corresponding to an eigenstate 
$e^{i({\pi}/{2})X}\,|j,m\rangle$ associated with $\mu=+i{\pi}/{2}$:
\begin{equation}
\left( i Y\right) \,e^{i({\pi}/{2})X}\,|j,m\rangle=	
	-i\,(m+{1}/{2} )\,e^{i({\pi}/{2})X}\,  |j,m\rangle 
\label{eq:lastar}
\end{equation}

Since $H_1=\G Y$, we therewith have complex eigenvalues for the Hamiltonian $H=
H_0+H_1$, and the eigenvectors of $H$ are in fact Gamow vectors (belonging to 
$\F^\times$~\cite{dirackets,bohm1,bohm2,gadella}) which exhibit pure exponential growth or
decay, depending on the sign of $\pm m$:
\begin{equation}
\psi^{G\;\pm}_{j,m}(t)=e^{\mp i({\pi}/{2})X}\,|j,m\rangle \,
	e^{-2i\Om jt\pm (\G /2)(2m+1)t}
\label{eq:gamow}
\end{equation}
Because $iY\sim H_1$ is in the familiar form of
a symmetric (hermitean) operator on Hilbert space, 
$e^{\pm i(\pi /2)X}\,|j,m \rangle$ cannot belong to the Hilbert space 
$\HH$ domain of the hermitean operator $H_1$, as proven by these complex 
eigenvalues of the hermitean $H_1 = iY$.  (One of the lessons of the
main body of this paper is that these Gamow vectors are associated
with semigroups of time evolution, with restricted time domains of
definition.  Thus, $\psi^{G+}_{j,m}\equiv \ts^G$ is defined only for
$t\le 0$ and $\psi^{G-}_{j,m}\equiv \psi^G$ is defined only for $t\ge
0$.) 

 These complex eigenvalues are 
totally unacceptable in the real algebra of a real group such as $SU(1,1)$, or 
in a representation of the same.  This is because $i \, \la \II \notin 
\mathfrak{su}(1,1)$, since the real algebra (and its associated group) 
is defined over the field of real scalars only.


\begin{thebibliography}{00}
\bibitem{note1} 
{Note the use of the symplectic form on
$\F\times\F^\times$ and $\HH\times\HH^\times$ is associated with a
scalar product.  In the usual physical notation for the scalar
product, $(\bullet \vert \bullet )$ or $\langle
\bullet\vert\bullet\rangle$, the ket $\vert\bullet\rangle$ would be
identified with the space and the bra $\langle \bullet \vert$
identified with the dual space.  Hence, the scalar product of 
physics is identified with $\FT\times\F$, and so involves the 
transpose of the symplectic form adopted by mathematicians.  
We shall blissfully ignore this distinction, since our physical 
concern is with the magnitude of the resulting scalar product.
Note that when working with dual pairs of spaces, $(\chi
,\chi^\times )$, many things are freed from dependence on the
particular choice of topology on $\chi$, meaning that many of the
properties we will deal with by proceeding in this way are
topologically invariant.  See, e.g., \cite{narici}. Thus, the
choice of seminorm topology for $\F$ or the 
choice of the norm topology for $\HH$ may actually have less effect on
the final results than one might at first suspect.  The use of the RHS
formalism for the description of resonances has the advantage of a
transparent mathematical rigor, whereas alternative treatments--such
as using  
Hilbert space operators outside of their apparent domain of
definition--often seem lacking in justification even if mathematical
care is 
used.  It is probable that careful examination of the RHS formalism
would lead justification of many hueristic devices used to describe 
resonances over the years.}
\bibitem{note2} 
{The essential purpose of making an alternative choice to the 
field $\C$ is the need for a geometric structure (in numerous places) 
which reflects the structure of the complex plane, and the requirement 
for a real algebra structure in order to have well defined adjoints,
as will be discussed below.  The commutative
real algebra $\C (1,i)\equiv \R\oplus i\circ \R$ 
has such a structure.  The algebra
$\C (1,i)$ is obtained by adding further structure to the field $\C$,
in order to obtain an algebraic representation of the complex plane.}
\bibitem{note3}  
{Since we use operators for which the
notion of hermiticity is not governing, our dynamical analogue of 
unitarity is defined using
the dual rather than the hermitean conjugate: that transformation $T$
have a symplectic action on a space requires that $T^\times T = TT^\times
=\II$~\cite{porteous}, which replaces for us the familiar rule in
Hilbert space 
$U^\dagger U =UU^\dagger =\II$.  Because we work with spaces having a
complex symplectic structure, this form of adjoint transformation
insures the transformations have a symplectic (=dynamical) action on
our spaces of states, i.e., $T$ so defined is a dynamical
transformation on $\F$ of $\rhs$.}
\bibitem{note16}  
{This procedure can be given the physical interpretation of 
representing the preparation procedure of an experiment 
as a combination of continuous canonical transformations, and the
continuation  
along similar lines to represent the decay and measurement processes 
as canonical 
transformations is straightforward.  There could ultimately be 
free oscillators at the outgoing end of the process.  The  
equation (\ref{eq:Htransform}) can be interpreted as the (active)
canonical  
transform, relating to the measurement process, or as merely setting
up a  formal evaluation of the constant $\G$ which characterizes the
random decay process.}
\bibitem{note18} 
{This means there is really no exponential catastrophe for the
Gamow vectors when the scalar product as a ``length function'' 
is properly defined.  The ``alternative normalization that always
seems to work'' of`\cite{kukulin}, is in fact
mathematically principled and proper.  Note this also involves
identification of the scalar product conjugate of a
Gamow vector with the Schwartz reflection principle conjugate.}
\bibitem{note20}   
{It will be seen that the
uniqueness requirement for the adjoint involution of our complex Lie 
algebra/group representation means we must work with multicomponent
vectors which are in fact spinors.  This issue
will be addressed in the fourth installment of this series~\cite{IV}.}
\bibitem{note21}  
{The fact that these Gamow vectors make up a set of zero measure has
physical significance for the ergodicity discussions in the third
installment of this series~\cite{III}.  If they had formed a set of positive
measure, they and the resonances they represent could not be
associated with entropy growth, for instance.  See, e.g.,~\cite{krylov}.}
\bibitem{note21x5} 
{We show in~\cite{note18} that there is no exponential catastrophe for 
properly defined Gamow states. There is
a scalar product well defined between the Gamow state arising during
$t\le 0$,  $|{\tilde\psi}^G\rangle$ and the Gamow state decaying
during $t\ge 0$, $|\Psi^G\rangle$.  This is an instance of the
Schwartz reflection principle, and is well defined because  
of the reflexitivity of the spaces $\Phi$ and $\Phi^\times$--one may
add a countable set of $|{\tilde\psi}^G\rangle$  
to $\Phi$ without adverse consequence, and this is what has, in
effect, occured.  This takes us out of the domain of strictly
infinitesimally generated semigroups, which we should point out here,
since our immediate concern is precisely 
the strictly infinitesimally generated semigroups, and no more.}
\end{thebibliography}
\end{document}